\documentclass[10pt]{wlscirep}

\usepackage{multirow}
\usepackage{array}
\usepackage{threeparttable}

\title{Continuous-variable source-device-independent quantum key distribution against general attacks}

\author[1]{Yichen Zhang}
\author[2]{Ziyang Chen}
\author[3]{Christian Weedbrook}
\author[1,*]{Song Yu}
\author[2]{Hong Guo}
\affil[1]{State Key Laboratory of Information Photonics and Optical Communications, Beijing University of Posts and Telecommunications, Beijing 100876, China}
\affil[2]{State Key Laboratory of Advanced Optical Communication Systems and Networks, Department of Electronics, and Center for Quantum Information Technology, Peking University, Beijing 100871, China}
\affil[3]{Xanadu, 372 Richmond St W, Toronto, M5V 2L7, Canada}
\affil[*]{yusong@bupt.edu.cn}

\begin{abstract}
The continuous-variable quantum key distribution with entanglement in the middle, a semi-device-independent protocol, places the source at the untrusted third party between Alice and Bob, and thus has the advantage of high levels of security with the purpose of eliminating the assumptions about the source device. However, previous works considered the collective-attack analysis, which inevitably assumes that the states of the source has an identical and independently distributed (i.i.d) structure, and limits the application of the protocol. To solve this problem, we modify the original protocol by exploiting an energy test to monitor the potential high energy attacks an adversary may use. Our analysis removes the assumptions of the light source and the modified protocol can therefore be called source-device-independent protocol. Moreover, we analyze the security of the continuous-variable source-device-independent quantum key distribution protocol with a homodyne-homodyne structure against general coherent attacks by adapting a state-independent entropic uncertainty relation. The simulation results indicate that, in the universal composable security framework, the protocol can still achieve high key rates against coherent attacks under the condition of achievable block lengths.
\end{abstract}
\begin{document}

\flushbottom
\maketitle
\thispagestyle{empty}

\section*{Introduction}

Quantum key distribution (QKD)~\cite{RevModPhys.74.145, RevModPhys.81.1301,Pirandola_RevModPhys_2019}, as one of the most practical quantum cryptography technology, allows two users (traditionally called Alice and Bob) to establish a set of secret keys exploiting both quantum mechanics and classical post-processing methods. This can provide information-theoretic security even against existing potential eavesdroppers.

Continuous-variable (CV) QKD protocol~\cite{RevModPhys.84.621, Entropy.17.6072}, of which the characteristic is that the information is encoded on the quadratures of the light field and measured with coherent measurement methods, e.g., homodyne~\cite{Phys.Rev.Lett.88.057902.2002} and heterodyne detection~\cite{Phys.Rev.Lett.93.170504.2004}, has developed rapidly. There are two main reasons resulting in CV-QKD attracting so much attention in recent years: it can be easily implemented with standard telecom components~\cite{Nature.Photon.13.839.2019,Zhang_Arxiv_2020} and compatible with wavelength division multiplexing~\cite{Photon.Tech.Lett.30.650.2018,Communications.Physics.2.9.2019}, and it can achieve high key rate in metropolitan distance~\cite{field_test}, which has advantages of short-range implementation.

There are plenty of CV-QKD protocols proposed to deal with different scenarios. In the case of fully trusted-device protocols, it is always assumed that both Alice and Bob are honest, and Eve can only control the quantum channels rather than the devices at the two parties. A large number of distinctive trusted-device protocols, including discrete modulation CV-QKD protocols~\cite{Phys.Rev.Lett.102.180504.2009, Phys.Rev.A.83.042312.2011, arXiv.1805.04249}, two-way protocols~\cite{Nat.Phys.4.726.2008, Int.J.Quantum.Inf.10.1250059.2012, J.Phys.B.47.035501.2014, Phys.Rev.A.92.062323.2015, Sci.Rep.6.22225.2016, J.Phys.B.At.Mol.Opt.Phys.50.035501.2017} and so forth, have been put forward to enrich the protocol design. However, because of the imperfection of the practical source and detection devices, a QKD system may be attacked by a potential eavesdropper, and it compromises the security of a protocol~\cite{New.J.Phys.20.103016.2018}. To eliminate all the loopholes of devices, fully device-independent protocols are proposed, which do not make any assumptions for all experimental devices and allows Eve to control them all. Nevertheless, those protocols face experimental challenges because they have to perform a detection-loophole-free Bell test~\cite{Phys.Rev.Lett.120.040406.2018}.

As a compromise, semi-device-independent (semi-DI) protocols are proposed, such as measurement-device-independent (MDI)~\cite{Nat.Photon.9.397.2015,Phys.RevA.89.052301, Phys.RevA.90.052325} and one-sided device-independent (1sDI)~\cite{Nat.Commun.6.8795.2015, Optica.3.634.2016} QKD protocols, to give a trade off between the security of some devices and the performance of a protocol, which regard that part of the protocol is honest and the other part is untrusted. Remarkably, both CV-MDI~\cite{LS.JW4A.33, Phys.Rev.A.97.052327.2018, Phys.Rev.A.98.012314.2018} and CV-1sDI protocols~\cite{Nat.Commun.6.8795.2015, Phys.Rev.Lett.109.100502.2012, Phys.Rev.A.90.042325.2014} have been analyzed against general coherent attacks, which improves the security analysis of the protocols.

CV-QKD with entanglement in the middle~\cite{Phys.Rev.A.87.022308.2013} is the protocol of which the source is placed at the untrusted third party in the middle and controlled by the malicious eavesdropper. Alice and Bob then measure one of the modes they received separately, with either homodyne or heterodyne detection. The goal of the protocol is that we do not need to give assumptions on the source, which is sometimes ill-characterised and unsafe in communication. Nevertheless, the security analysis of the CV-QKD with entanglement in the middle protocol is only confined to the collective attack cases, which inevitably assumes that the states of the source has identical and independently distributed (i.i.d) structure, i.e., ${\rho _{{A^n}{B^n}}} = \sigma _{AB}^{ \otimes n}$, leading to the protocol unable to reach the original idea of source-device-independent (SDI).

Inspired by the security analysis technique used in the 1sDI protocol by F. Furrer \emph{et al.}~\cite{Phys.Rev.Lett.109.100502.2012, Phys.Rev.A.90.042325.2014}, we adapt one type of state-independent entropic uncertainty relation with CVs to analyse the security of the CV-QKD with entanglement in the middle protocol under coherent attacks and only consider the case that both Alice and Bob perform homodyne detections. We modify the original protocol by exploiting an energy test at the reconciliation side (Bob's side for reverse reconciliation as an example) to monitor the potential high energy attacks an adversary may use. By properly quantifying the correlation between Alice's and Bob's data, which could be used for estimating Eve's knowledge of the raw key, we obtain the secret key rate of a finite number of exchanged signals supposing that the strategy Eve exploits is a coherent attack. Our analysis removes the assumptions of the light source and assumes that the sampling process performed in Alice's and Bob's sides are i.i.d, which is needed for exploiting the entropic uncertainty relation. Therefore, The modified protocol can be called CV-SDI QKD protocol. Finally, simulation shows that even when the coherent attack is considered, CV-QKD with entanglement in the middle can still reach a non-zero key rate over short distance, without giving any constrains of the source.

\section*{Results}

\textbf{The original CV-QKD protocol with entanglement in the middle against collective attacks}

\begin{figure}[t]
	\centering
	\includegraphics[width=13.35 cm]{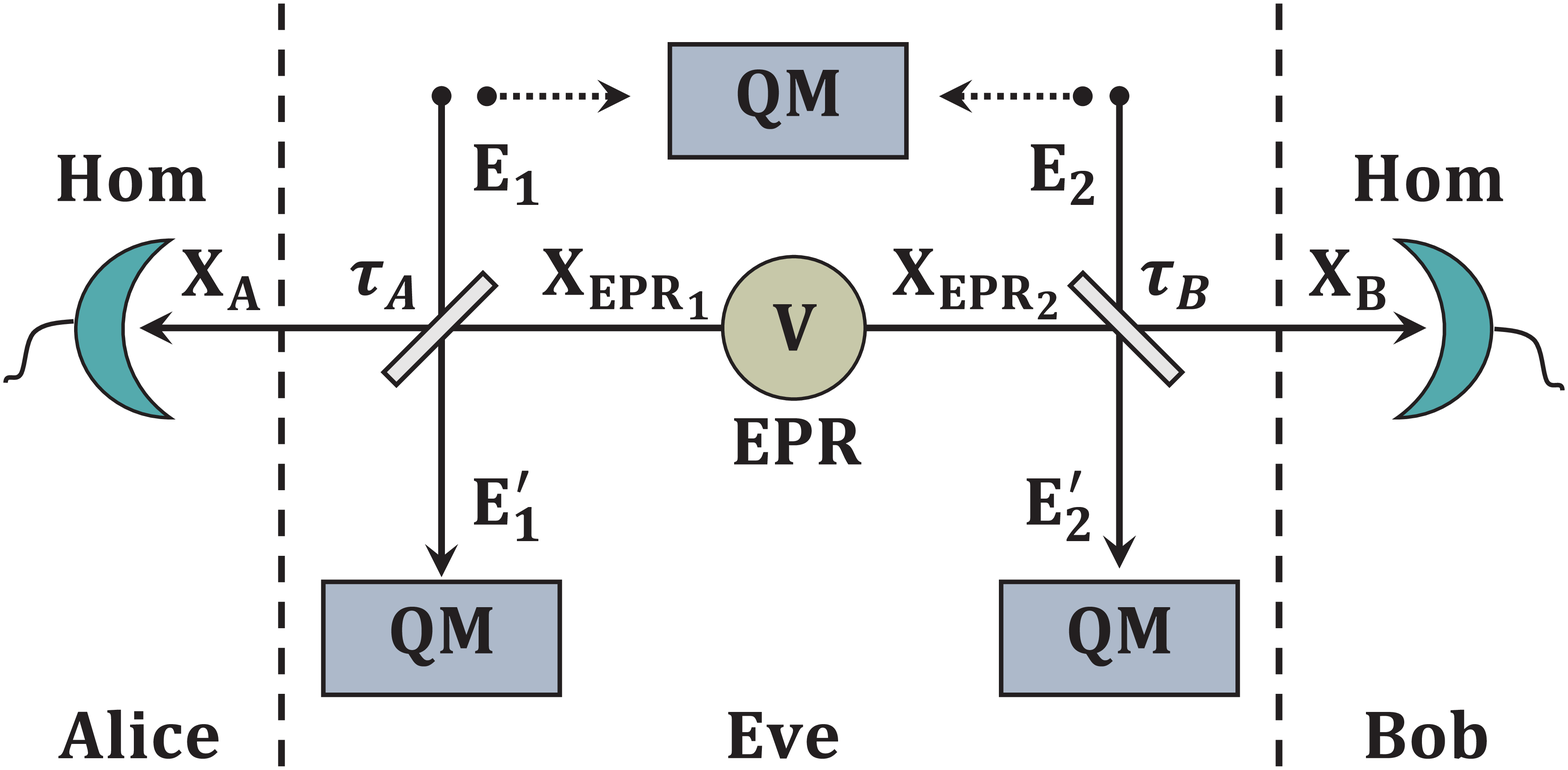}\\
	\caption{Schematic of the entanglement-in-the-middle CV-QKD protocol~\cite{Phys.Rev.A.87.022308.2013}. EPR: untrusted two-mode squeezed state with variance $V$. Hom: homodyne detection. QM: quantum memory. Only the homodye detections are discussed here and Eve's attacks are considered as two correlated modes attacks without loss of generality.}
	\label{entangling_in_the_middle}
\end{figure}

We begin by describing the CV-QKD protocol with entanglement in the middle, which was originally proposed in Ref.~\cite{Phys.Rev.A.87.022308.2013}. A two-mode squeezed vacuum state EPR, with an unknown variance $V$, is prepared by the untrusted third party, see Fig.~\ref{entangling_in_the_middle}. The EPR source can be created either by an untrusted communication party Charlie or by the potential adversary Eve. The two modes of an EPR source, e.g., EPR$_1$ and EPR$_2$, are sent to Alice and Bob separately through quantum channels. As the general assumption in QKD is that both of the two quantum channels could be totally controlled by potential eavesdropper Eve; leading to the introduction of loss and noise to the states after transmission. Assuming the quadratures of the two modes of the EPR source are ${\hat X_{EP{R_{\rm{1}}}}}$ and ${\hat X_{EP{R_{\rm{2}}}}}$ with the covariance matrix (CM)

\begin{equation}\
{\gamma _{EPR}{\rm{ = }}\left( {\begin{array}{*{20}{c}}
		{V\textbf{I}}&{\sqrt {{V^2} - 1} \textbf{Z}}\\
		{\sqrt {{V^2} - 1} \textbf{Z}}&{V\textbf{I}}
		\end{array}} \right)},
\end{equation}
where $\textbf{I} = \rm{diag[1,1]}$ and $\textbf{Z} = \rm{diag[1,-1]}$, and the transmissivities of two channels are ${\tau _A}$ and ${\tau _B}$ respectively, then we have the quadratures after transmissions, given by
\begin{align}\
{\hat X_{\rm{A}}}{\rm{ = }}\sqrt {{\tau _A}} {\hat X_{EP{R_1}}}{\rm{ + }}\sqrt {{\rm{1 - }}{\tau _A}} {\hat X_{{E_1}}},                   \notag\\
{\hat X_{\rm{B}}}{\rm{ = }}\sqrt {{\tau _B}} {\hat X_{EP{R_2}}}{\rm{ + }}\sqrt {{\rm{1 - }}{\tau _B}} {\hat X_{{E_2}}},
\end{align}\
where $E_{1}$ and $E_{2}$ are the ancillary systems which Eve inject into the links to attack the protocol. The two-correlated-mode eavesdropping strategy is considered here, which is the general two-mode attack strategy, where the CM ${\gamma _{{E_1}{E_2}}}$ of the two correlated modes is
\begin{equation}\
{\gamma _{{E_1}{E_2}}} = \left( {\begin{array}{*{20}{c}}
	{{\omega _A}\rm{\textbf{I}}}&\rm{\textbf{G}}\\
	\rm{\textbf{G}}&{{\omega _B}\rm{\textbf{I}}}
	\end{array}} \right),
\end{equation}
where ${\omega _A}$ and ${\omega _B}$ are the variance of modes $\rm{{E_1}}$ and $\rm{{E_2}}$, and the correlation term $\rm{\textbf{G}}= \rm{diag}\left[ {g,g'} \right]$ with the correlation parameters $g$ and $g'$ satisfying the constraints shown in Ref.~\cite{New.J.Phys.15.113046.2013}. The attack is optimal by setting modes $\rm{{E_1}}$ and $\rm{{E_2}}$ as coherent given in Refs.~\cite{Nat.Photon.9.397.2015, Phys.Rev.A.91.022320.2015, arXiv:1509.04144}.

Originally, Alice and Bob perform quadrature measurements via homodyne or heterodyne detections, and in this paper, we only consider the scenario that both Alice and Bob employ homodyne detections to get one measurement result, i.e., quadrature $x$ or $p$. After finishing the state preparation and measurement phases, both Alice and Bob announce which quadrature they choose through an authenticated pubic channel to sift their keys. They hold the data for which the selected quadratures are the same and discard the rest. Finally, the two communication parties proceed with classical data post-processing, namely parameter estimation, error correction and privacy amplification to distill their keys.

In the case of collective attacks setting, the state ${\rho _{{A^N}{B^N}{E^N}}}$ after all runs can be considered as a tensor product state, namely ${\rho _{{A^N}{B^N}{E^N}}} = \rho _{ABE}^{ \otimes N}$, where $N$ is the total number of quantum signals exchanged during the protocol. In this paper, we only focus on the asymptotic case under collective attacks to show the ideal performance of the protocol, where the total number of quantum states $N$ tends to infinite. The asymptotic secret key rate $K_{collective}^{asym}$ (for reverse reconciliation) is given by the Devetak-Winter formula~\cite{Proc.Roy.Soc.A.461.207.2005}, which reads
\begin{equation}\
K_{collective}^{asym} = \max \left\{ {\beta I\left( {A:B} \right) - \chi \left( {B:E} \right),0} \right\},
\label{D_W_relation}
\end{equation}
where $\beta$ is the reconciliation efficiency, $I\left( {A:B} \right)$ is the classical mutual information between Alice's and Bob's data, and $\chi \left( {B:E} \right)$ is the Holevo information between Bob's data and the eavesdropper \cite{Probl.Inf.Transm.9.177.1973}. This is given by $\chi \left( {B:E} \right) = S\left( E \right) - S\left( {E|B} \right)$, where $S\left( E \right)$ is the von Neumann entropy of Eve and $S\left( {E|B} \right)$ is the conditional von Neumann entropy of Eve given Bob's information.

$\chi \left( {B:E} \right)$ can be bounded with the help of the Gaussian state extremality theorem~\cite{Phys.Rev.Lett.97.190503.2006, Phys.Rev.Lett.96.080502.2006} in the case of collective attacks, hence we assume that the state ${\rho _{AB}}$ is Gaussian to minimize the final secret key rates, which can be calculated from the CM. A detailed derivation of the CM and the key rate can be seen in Methods section.

\textbf{The modified CV-SDI QKD protocol against general coherent attacks}

In the case of general coherent attacks, the assumption that ${\rho _{{A^N}{B^N}{E^N}}}$ has tensor product structure is invalid, so we cannot apply Eq.~(\ref{D_W_relation}) directly to bound the security key rate after finite runs of the protocol. There are in general two main security-proof techniques developed in CV-QKD to handle coherent attack issues. One method is the de Finetti theorem~\cite{Phys.Rev.Lett.114.070501.2015, Phys.Rev.Lett.118.200501.2017}, which have the ability to reduce the security from coherent attacks to collective attacks, and it was successfully employed to analyse the protocol which has some symmetric properties~\cite{Phys.Rev.A.97.052327.2018}. The alternative is the entropic uncertainty relation~\cite{Phys.Rev.Lett.109.100502.2012, Phys.Rev.A.90.042325.2014,Entropy.21.652.2019}, which requires that the protocol needs to randomly measure between two quadratures and perform the sifting process~\cite{Nat.Commun.6.8795.2015, Phys.Rev.A.98.012314.2018}. We exploit the latter tool in this paper to obtain the security of the entanglement-in-the-middle protocol with homodyne-homodyne structure against coherent attacks. We point out that the protocol in Ref.~\cite{Phys.Rev.A.90.042325.2014} has no assumption on Alice's side (also be treated as the source side), thus it is also called one-sided device independent protocol. In our protocol, there is also no assumption on the source. However, since the structure of our protocol is a network structure, where the source is located in a third party, and Alice and Bob only perform measurements, this structure is very different from previous protocol, where the source is located in one side of the protocol. We named our protocol ``source device independent'' to distinguish it from previous one-sided device independent protocols.

We analyse the protocol under general coherent attacks with untrusted source in the middle by adapting the approach described in Ref.~\cite{Phys.Rev.A.90.042325.2014}. Thanks to the composable security framework, we have the ability to study the protocol considering some imperfect situation, such as the practical detection model, the energy test and finite-size effect, which allows us to modify the protocol in coherent-attack case.

\textbf{Simulation}

\begin{figure}[h]
	\centering
	\includegraphics[width=10 cm]{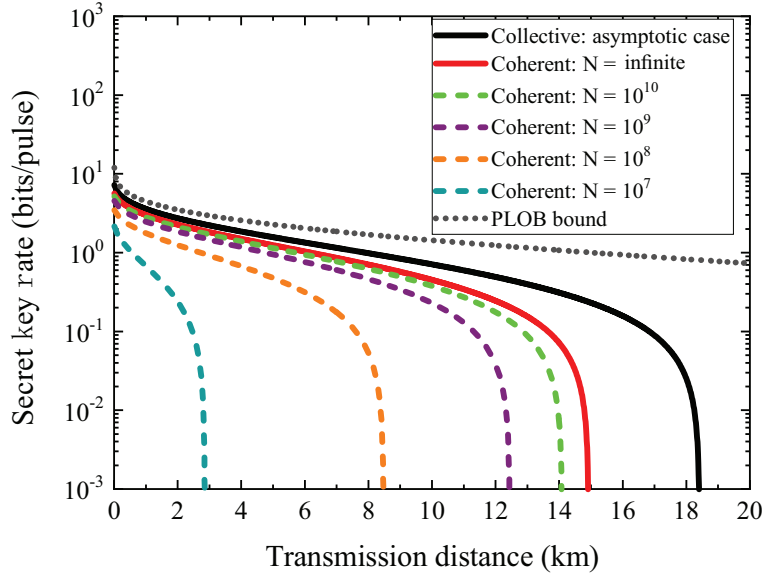}\\
	\caption{Secret key rates of the CV-SDI QKD protocol. The protocol is under symmetric configuration with ${\tau _A} = {\tau _B} = \sqrt {{T}}$ where $T$ is the total transmissivity of the channel. We consider the protocol with perfect reconciliation efficiency $\beta  = 1$ and ideal modulation variance $V = {10^5}$. We also set the excess noise as $\xi  = 0.001$ in each channel and the overall security parameter is smaller than ${10^{ - 20}}$. The gray dot line is the PLOB bound~\cite{Nat.Commun.8.15043.2017} and the black solid line is the key rate under collective attacks. The red solid line is the key rate under coherent attacks with infinite exchanged signals. The four dashed lines, from top to bottom, are the secret key rates under coherent attacks, with the block lengths from ${10^{10}}$ to ${10^{7}}$.}
	\label{simulation_key_rates}
\end{figure}

\begin{figure}[h]
	\centering
	\includegraphics[width=10 cm]{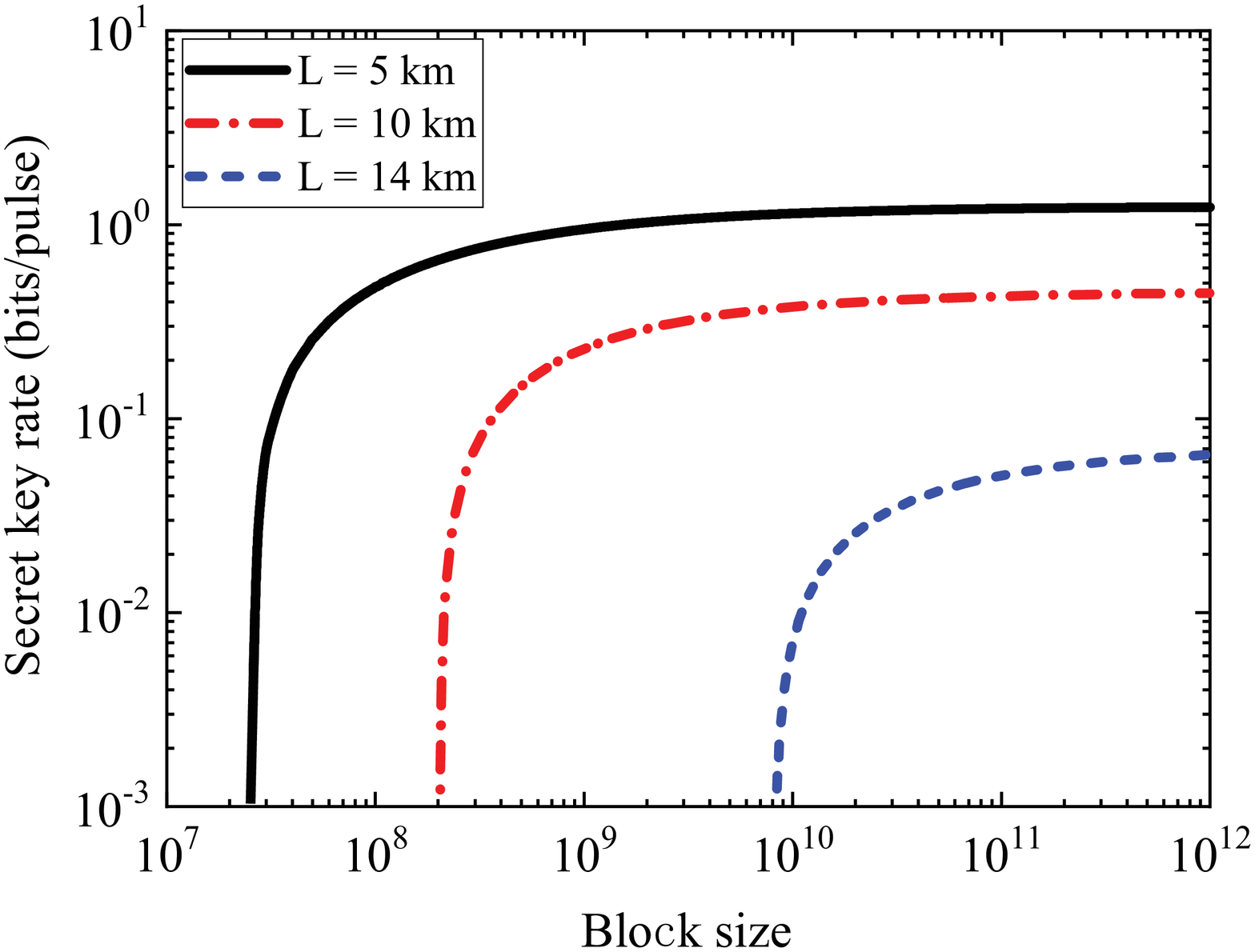}\\
	\caption{Secret key rates as functions of block size of the CV-SDI QKD protocol. The black solid line shows the performance with the distance of 5km. The red dot-dashed line and the blue dashed line are the key rates of the protocol with distances of 10km and 14km, respectively. The other parameters are as in Fig.~\ref{simulation_key_rates}}
	\label{N_change}
\end{figure}

Using the results in the previous section, we can plot the secret key rate as a function of the total transmission distance focusing on the symmetric configuration where we set ${\tau _A} = {\tau _B} = \sqrt {{T}}$ and $T$ is the transmissivity of the channel. The simulations are under two-mode optimal attacks to show the performance of the protocol and both collective and coherent attack scenarios are discussed shown in Fig.~\ref{simulation_key_rates}. Note that modeling an eavesdropper's attack behavior here does not limit the eavesdropping ability, but just for the convenience of simulations. Actually, in experiment, we only need to know the parameter estimation data $\left\{ {x_A^{pe},p_A^{pe}} \right\}$ and $\left\{ {x_B^{pe},p_B^{pe}} \right\}$ of Alice and Bob to execute the security analysis of the protocol. Therefore there is no need to assume which model Eve's attack strategy belongs to before the protocol starts. Modeling attacks of eavesdroppers with two-mode coherent attacks yields the worst performance of the protocol~\cite{Nat.Photon.9.397.2015, Phys.Rev.A.91.022320.2015, arXiv:1509.04144}, thus we use this modeling method to well reflect the performance of the protocol. The results are shown in Fig.~\ref{simulation_key_rates} and Fig.~\ref{N_change}, where Fig.~\ref{simulation_key_rates} shows the secret key rates of the CV-SDI QKD as the function of transmission distance under different block sizes, while Fig.~\ref{N_change} is the key rate varying with the block size.

\section*{Discussion}

In order to facilitate the analysis of the performance of the protocol, we simulate the key rate with some ideal parameters. For instance, we assume that the protocol has an ideal modulation variance $V = {10^5}$ (which could replace an infinite modulation variance) and perfect reconciliation efficiency $\beta  = 1$. Also, to get the lower bound of the protocol, we set $g = \min \left\{ {\sqrt {\left( {{\omega _A} - 1} \right)\left( {{\omega _B} + 1} \right)} ,\sqrt {\left( {{\omega _A} + 1} \right)\left( {{\omega _B} - 1} \right)} } \right\}$ and ${\omega _A} = {\omega _B} = 1 + {{T\xi } \mathord{\left/
		{\vphantom {{T\xi } {\left( {1 - T} \right)}}} \right. \kern-\nulldelimiterspace} {\left( {1 - T} \right)}}$ with excess noise $\xi$ in one channel for two-mode optimal attacks. In the coherent attack cases, we set the interval parameter $\alpha$ to $52$~\cite{Phys.Rev.Lett.109.100502.2012} and the overall security parameter is smaller than ${10^{ - 20}}$. Meanwhile, the parameter ${M_{th}}$ is set to $12$ to ensure that the energy test fails with probability smaller than ${10^{ - 20}}$.

In Fig.~\ref{simulation_key_rates}, the gray dot line shows the Pirandola-Laurenza-Ottaviani-Banchi (PLOB) bound~\cite{Nat.Commun.8.15043.2017}, which gives an upper bound of the secret key capacity of the lossy channel. The black solid line is the asymptotic key rate under collective attacks, and the longest transmission distance is over 18km, which is a little shorter than that of the two-mode individual attacks case (where the correlation parameter $g=0$)~\cite{Phys.Rev.A.87.022308.2013}. The other five curves, from top to bottom, describe the key rates under coherent attacks. The red solid curve is obtained for $N \to \infty$, and the other dashed lines describe the rate for $N = {10^{10}}$ to $N = {10^{7}}$ with finite exchanged signals. In Fig.~\ref{N_change}, we also plot the secret key rate under coherent attacks as a function of block size for different distances. The distances are 5km, 10km and 14km, respectively. We point out that when the block size reduces, the secret key rate decreases, and it is not achievable if the block size is below ${10^7}$.

We notice that there is a gap between the performance of CV-QKD protocol with entanglement in the middle under collective attacks and that under asymptotic coherent attacks cases. The reason is that the bound given by the entropic uncertainty relation is not very tight especially in the high losses regime, which has been shown in Ref.~\cite{Phys.Rev.A.90.042325.2014}.

In conclusion, we have analyzed the security of continuous-variable source-device-independent quantum key distribution protocol against general coherent attacks, where the source of the protocol is untrusted and may be controlled by the malicious adversary. By exploiting the state-independent entropic uncertainty relation together with the energy test, our analysis has no assumptions on the source, making the protocol source-device-independent even under coherent attacks. The simulation results indicate that, in the universal composable security framework, the protocol is still secure, achieving high key rates against coherent attacks under the condition of achievable block lengths ($N$ from ${10^7}$ to ${10^{10}}$).

\section*{Methods}

\textbf{Covariance matrix and the secret key rate under collective attacks}

The final bipartite quantum state ${\rho _{AB}}$ of Alice and Bob has the CM with the form
\begin{equation}\
{\gamma _{AB}} = \left( {\begin{array}{*{20}{c}}
	{a\rm{\textbf{I}}}&{c\rm{\textbf{Z}}}\\
	{c\rm{\textbf{Z}}}&{b\rm{\textbf{I}}}
	\end{array}} \right),
\end{equation}
where
\begin{align}
&a = {\tau _A}V + \left( {1 - {\tau _A}} \right){\omega _A},     \notag\\
&b = {\tau _B}V + \left( {1 - {\tau _B}} \right){\omega _B},     \notag\\
&c = \sqrt {{\tau _A}{\tau _B}} \sqrt {{V^2} - 1}  - g\sqrt {1 - {\tau _A}} \sqrt {1 - {\tau _B}},
\end{align}
and we let
\begin{equation}\
g = \min \left\{ {\sqrt {\left( {{\omega _A} - 1} \right)\left( {{\omega _B} + 1} \right)} ,\sqrt {\left( {{\omega _A} + 1} \right)\left( {{\omega _B} - 1} \right)} } \right\}
\end{equation}
by setting modes ${E_1}$ and ${E_2}$ are coherent. Then the secret key rate $K_{collective}^{asym}$ can be calculated by Eq. (\ref{D_W_relation}) if we restrict our discussion in reverse reconciliation cases. The mutual information between Alice's and Bob's data can be described as
\begin{equation}\
I\left( {A:B} \right) = \frac{1}{2}{\log _2}\left( {\frac{a}{{a - {{{c^2}} \mathord{\left/
					{\vphantom {{{c^2}} b}} \right.
					\kern-\nulldelimiterspace} b}}}} \right).
\end{equation}
To obtain the von Neumann entropy $S\left( E \right)$ and $S\left( {E|B} \right)$, we always assume that Eve can purify the whole system in order to maximize her information, thus we have $S\left( E \right) = S\left( {AB} \right)$ and $S\left( {E|B} \right) = S\left( {A|B} \right)$. ${S\left( {AB} \right)}$ is a function of the symplectic eigenvalues ${\lambda _{1,2}}$ of ${\gamma _{AB}}$, which reads
\begin{equation}\
S\left( {AB} \right) = G\left[ {{{\left( {{\lambda _1} - 1} \right)} \mathord{\left/
			{\vphantom {{\left( {{\lambda _1} - 1} \right)} 2}} \right.
			\kern-\nulldelimiterspace} 2}} \right] + G\left[ {{{\left( {{\lambda _2} - 1} \right)} \mathord{\left/
			{\vphantom {{\left( {{\lambda _2} - 1} \right)} 2}} \right.
			\kern-\nulldelimiterspace} 2}} \right],
\end{equation}
where
\begin{equation}\
G\left( x \right) = \left( {x + 1} \right){\log _2}\left( {x + 1} \right) - x\log x,
\end{equation}
and
\begin{equation}\
\lambda _{1,2}^2 = \frac{1}{2}\left[ {\Delta  \pm \sqrt {{\Delta ^2} - 4{D^2}} } \right],
\end{equation}
where we use the notations that $\Delta  = {a^2} + {b^2} - 2{c^2}$ and $D = ab - {c^2}$. After Bob performs homodyne detection, Alice's CM conditioned on Bob's measurement results will transform to
\begin{equation}\
\gamma _A^{{x_b}} = {\gamma _A} - \Sigma _C^T{\left( {X{\gamma _B}X} \right)^{ - 1}}{\Sigma _C},
\end{equation}
where ${\gamma _A} = a\rm{\textbf{I}}$, ${\gamma _B} = b\rm{\textbf{I}}$, ${\Sigma _C} = c\rm{\textbf{Z}}$ and $X = \left[ {1,0;0,0} \right]$. $S\left( {A|B} \right) = G\left[ {{{\left( {{\lambda _3} - 1} \right)} \mathord{\left/
			{\vphantom {{\left( {{\lambda _3} - 1} \right)} 2}} \right. \kern-\nulldelimiterspace} 2}} \right]$ is a function of the symplectic
eigenvalue ${{\lambda _3}}$ of the covariance matrix $\gamma _A^{{x_b}}$ with ${\lambda _3} = \sqrt {a\left( {a - {{{c^2}} \mathord{\left/
				{\vphantom {{{c^2}} b}} \right.\kern-\nulldelimiterspace} b}} \right)}$. Therefore, the secret key rate under collective attacks when the reverse reconciliation is performed is
\begin{equation}\
K_{collective}^{asym} = \beta I\left( {A:B} \right) - \left[ {S\left( {AB} \right) - S\left( {A|B} \right)} \right].
\end{equation}

\textbf{The practical detection model and the measurement phase}

We model the practical detector as an ideal homodyne detector followed by an analog-to-digital converter (ADC) with finite sampling range, and therefore the measurement process can be divided into two steps.

In Step 1, Alice and Bob exploit ideal homodyne detectors to measure the input signal with infinite ranges and resolutions. The measurement quadratures are ideal continuous variables with infinite dimensions, hence the measurement results are also continuous. Assuming that the sifting process is done, we denote the outputs of ideal homodyne detectors as ${Q_A}$ and ${Q_B}$ in two sides. In general CV-QKD scenario, the statistical distribution of each outcome should follow a Gaussian distribution.

In order to obtain a tight bound using the entropic uncertainty relation, we need to rescale one of two results, ${Q_A}$ or ${Q_B}$, and ensure that Alice's and Bob's measurement outcomes have high correlations after transmission through untrusted channels. We use the transformations below (using Alice as an example) to scale the quadrature measurements:
\begin{equation}\
{Q_A} \to {\tilde Q_A} = {t_q}{Q_A},
\label{rescale}
\end{equation}
where $t_q$ denotes the rescaling factor related to the channel losses of Alice and Bob, which can be determined by matching the variances of Alice's and Bob's measurement results. Supposing that $m$ signals are randomly chosen to do the parameter estimation, the average value of quadrature measurement results
both in Alice's and Bob's sides can be estimated by
\begin{equation}\
\hat E\left( {{Q_A}} \right) = \frac{1}{m}\sum\limits_{i = 1}^m {Q_A^i} ,\quad \hat E\left( {{Q_B}} \right) = \frac{1}{m}\sum\limits_{i = 1}^m {Q_B^i} ,
\end{equation}
where ${Q_A} = \left\{ {Q_A^i} \right\}_{i = 1}^m$ and ${Q_B} = \left\{ {Q_B^i} \right\}_{i = 1}^m$, and it is easy to estimate the parameter $t_q$ by \cite{Phys.Rev.A.98.012314.2018}
\begin{equation}\
{{\hat t_q}} = \sqrt {\frac{{\sum\limits_{i = 1}^m {{{\left( {Q_B^i - \hat E\left( {{Q_B}} \right)} \right)}^2}} }}{{\sum\limits_{i = 1}^m {{{\left( {Q_A^i - \hat E\left( {{Q_A}} \right)} \right)}^2}} }}}.
\end{equation}

In the symmetric case, where the channel losses and noises of Alice and Bob are approximately the same, we can simplify the analysis by assuming that ${t_q} \approx 1$.

In Step 2, the ADCs with finite range and finite precision followed by homodyne detectors are exploited to discretize continuous measuring intervals into discrete intervals, and the continuous variables ${\tilde Q_A}$ and ${Q_B}$ are also discretized. The measurement results are grouped into intervals:
\begin{equation}\
\left( { - \infty , - \alpha } \right], \left( { - \alpha   , - \alpha  + \delta } \right], ....,\left( {  \alpha  - \delta ,  \alpha  } \right], \left( {\alpha   ,\infty } \right),
\end{equation}
where $\alpha $ is the maximum discretization range of the ADCs, which takes the finite range of detectors into consideration, and $\delta$ denotes the resolution of the measurement, which shows how much detail the detector can detect. The corresponding outcome alphabet is denoted by $\chi  = \left\{ {1,2,...,{{2\alpha } \mathord{\left/{\vphantom {{2\alpha } \delta }} \right.\kern-\nulldelimiterspace} \delta }} \right\}$, where we assume ${{2\alpha } \mathord{\left/{\vphantom {{2\alpha } \delta }} \right.\kern-\nulldelimiterspace} \delta } \in \mathbb{N}$ and every measurement outcome corresponds to one of the intervals. After this step, the continuous outcomes are replaced by the discrete results, which are denoted by
\begin{equation}\
{\tilde Q_A}\mathop  \to \limits^{discrete} {X_A}, \quad {Q_B}\mathop  \to \limits^{discrete} {X_B}.
\end{equation}

This detection model can effectively illustrate the practical detector with finite range and resolution, without considering the efficiency of the detector, which could be modeled by a beam splitter with transmissivity $T_{d} $ \cite{Phys.Rev.Lett.102.130501.2009}. However, the ``discretization'' process may cause security issues when compared with the ideal detection case since the detection results are missing information about the quadratures. One issue is that any measurement outcomes inside one of the equal-length intervals $\left( { - \alpha   , - \alpha  + \delta } \right], ....,\left( {  \alpha  - \delta ,  \alpha  } \right]$ will map to the same value and it may cause a reduction in the information about the state within each sampling interval due to the finite sampling bits. This effect can be suppressed by increasing the number of sampling bits. The other problem is caused by two intervals with infinite length, namely $\left( { - \infty , - \alpha } \right]$ and $\left[ {\alpha ,\infty } \right)$, and users cannot know the full information about the state outside the detection range. In other word, users cannot distinguish whether the energy of the measured pulse is low or high, which may leave some loopholes for eavesdropping. This problem can be solved by the energy test solution.

\textbf{The energy test}

\begin{figure}[t]
	\centering
	\includegraphics[width=10 cm]{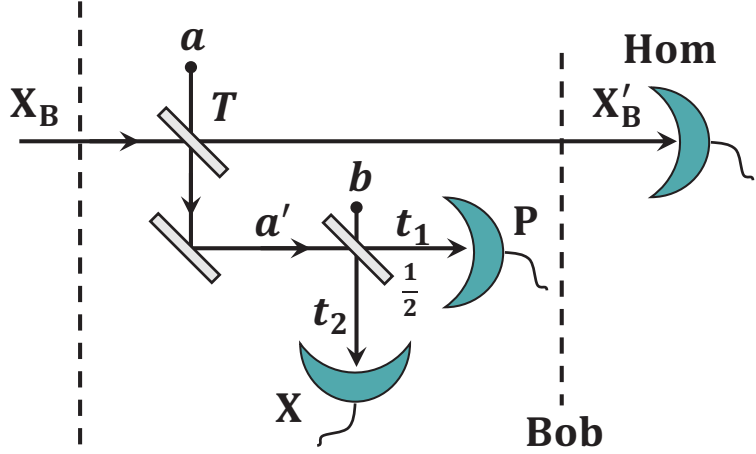}\\
	\caption{Schematic of the energy test at Bob's side. Bob uses a beam splitter with transmissivity $T$ to split the incoming signal into two parts. The transmission mode $\rm{{X'_B}}$ is used for generating Bob's data and the reflection mode $a'$ is exploited to perform the energy test. $a$ and $b$ are two vacuum modes induced by beam splitters. Modes ${t_1}$ and ${t_2}$ are the output modes of the balanced beam splitter used for checking whether $\left| {{q_{{t_1}}}} \right|$ and $\left| {{p_{{t_2}}}} \right|$ are below a certain threshold.}
	\label{energy_test}
\end{figure}

For fear of the large energy attack that Eve may exploit during the communication process, the protocol should be modified by adding the energy test step to ensure that the energy of measured states is below a certain threshold. We adapt the energy test method proposed in Ref.~\cite{Phys.Rev.A.90.042325.2014} to study entanglement-in-the-middle protocol to remove the assumption of the source in the security analysis, which should be considered in trusted source scenario~\cite{Phys.Rev.Lett.109.100502.2012}, hence this protocol also can be called \emph{source-device-independent} protocol.

Assuming that the protocol is performed with reverse reconciliation, the energy test is exploited in Bob's side before Bob performs the measurement step, which is described in Fig.~\ref{energy_test}. Bob uses a beam splitter with almost perfect transmissivity $T$ to split incoming mode $\rm{X_{B}}$ into two parts, and $a$ is the vacuum mode introduced by the other port of the beam splitter. Mode $\rm{{X'_B}}$ is the transmitted mode of the output used for generating Bob's raw data using a homodyne detector, and $a'$ is the reflected mode for the energy test. The reflected mode $a'$ is measured by a heterodyne detector, which consists of a balanced beam splitter and two homodyne detectors. Modes ${t_1}$ and ${t_2}$ are the output modes of the balanced beam splitter used for checking whether the amplitude of one output $\left| {{q_{{t_1}}}} \right|$ and the phase of the other output $\left| {{p_{{t_2}}}} \right|$ are below a certain threshold. If for every measured signal, both the amplitude $\left| {{q_{{t_1}}}} \right|$ and the phase $\left| {{p_{{t_2}}}} \right|$ are not larger than the threshold ${M_{th}}$, we say that the energy test passes; and the protocol aborts otherwise. The probability that Bob measures with homodyne detection larger than the detection range $\alpha$ can be bounded by the function $\Gamma \left( {\alpha,T,{M_{th}} } \right)$, which reads~\cite{Phys.Rev.A.90.042325.2014}
\begin{equation}\
\Gamma \!\left( {\alpha,T,{M_{th}} } \right): =\! \frac{{\sqrt {1 \!+\! \lambda } \! +\! \sqrt {1\! +\! {\lambda \!^{ - 1}}} }}{2}\exp \!\left( { - \frac{{{{\left( {\mu \alpha \! -\! {M_{th}}} \right)}^2}}}{{T\left( {1\! +\! \lambda } \right)/2}}} \right),
\end{equation}
where $\mu  = \sqrt {\frac{{1 - T}}{{2T}}}$ and $\lambda  = {\left( {\frac{{2T - 1}}{T}} \right)^2}$. The smoothness of the energy test ${\tilde \epsilon }$ further can be bounded by
\begin{equation}\
\tilde \epsilon  = \sqrt {\frac{{2n\Gamma \left( {\alpha,T,M_{th} } \right)}}{{{p_{pass}}}}}.
\label{energy_test_parameter}
\end{equation}

\textbf{Finite-size effect and the key rate}

In the coherent-attack scenario, due to the leftover hash lemma, the ${\epsilon _c}$-correct and ${\epsilon _s}$-secret key of length $\ell _{\sec }$ can be extracted~\cite{arXiv.0512258.2006}, which can be expressed by
\begin{equation}\
{\ell _{\sec }} \le H_{\min }^\epsilon {\left( {{X_B^{key}}|E} \right)_\rho } - {\ell _{EC}} - {\log _2}\frac{1}{{\epsilon _1^2{\epsilon _c}}} + 2,
\label{EUR_key_rata}
\end{equation}
where ${\ell _{EC}}$ denotes the leaked information in error correction step, and it can be estimated before the error correction begin during the parameter estimation phase, $H_{\min }^\epsilon \left(  {{X_B^{key}}|E} \right)$ is the smooth conditional min-entropy of data $X_B^{key}$ with smoothing parameter $\epsilon$, conditioned on the information Eve may have, which quantifies Eve's uncertainty about the Alice's measurement outcomes. $\epsilon$ satisfies $\epsilon  \le {{\left( {{\epsilon _s} - {\epsilon _{\rm{1}}}} \right)} \mathord{\left/{\vphantom {{\left( {{\epsilon _s} - {\epsilon _{\rm{1}}}} \right)} {2{p_{pass}} - 2\tilde \epsilon }}} \right. \kern-\nulldelimiterspace} {2{p_{pass}} - 2\tilde \epsilon }}$, where $p_{pass}$ is the probability that the parameter estimation step passes, ${\tilde \epsilon}$ is the security parameter related to the energy test given in Eq.~(\ref{energy_test_parameter}) and we choose ${\epsilon _1} = {{\epsilon _s} \mathord{\left/
		{\vphantom {e {\rm{2}}}} \right.\kern-\nulldelimiterspace} {\rm{2}}}$ for simplification~\cite{New.J.Phys.15.053049.2013}. Equation~(\ref{EUR_key_rata}) is a CV type key formula considering the quantum side information $E$ in infinite-dimensional Hilbert space~\cite{Phys.Rev.Lett.109.100502.2012}.

The parameter ${\ell _{EC}}$ can be easily obtained by publishing some of Bob's data (in the reverse reconciliation case), which is
\begin{equation}\
{\ell _{EC}} = H\left( {{X_B}} \right) - \beta I\left( {{X_B}:{X_A}} \right),
\end{equation}
where $H\left( {{X_B}} \right)$ denotes the discrete Shannon entropy of the data in Bob side, which can be described by
\begin{equation}
H\left( {{X_B}} \right) =  - \sum\limits_{i = 1}^n {p\left( {{x_i}} \right) {\log _2} p\left( {{x_i}} \right) - {\log _2} \delta },
\end{equation}
and $I\left( {{X_B}:{X_A}} \right)$ is the mutual information between Alice and Bob.

Our target is to bound the smooth min-entropy $H_{\min }^\epsilon \left(  {{X_B^{key}}|E} \right)$ in the presence of quantum adversaries. The entropic uncertainty relations were originally introduced in discrete variable QKD to bound the min-entropy and to show the protocols' security~\cite{Nat.Phys.6.659.2010, Phys.Rev.Lett.106.110506.2011}. They were thereafter extended to infinite dimensions by F.~Furrer \emph{et al.}~\cite{Commun.Math.Phys.306.165.2011, J.Math.Phys.55.122205.2014}. Therefore we exploit one type of uncertainty relation formula shown in Ref.~\cite{Phys.Rev.A.90.042325.2014} to bound the min-entropy in the entanglement-in-the-middle protocol, and the feature of the entropic uncertainty relation together with the energy test, resulting in making the protocol being source-device-independent.

Entropic uncertainty relation gives a bound of guessing the uncertainty that the eavesdropper may have, when both communication parties randomly measure in two bases. The relationship between smooth min- and max- entropies is given by
\begin{equation}\
H_{\min }^\epsilon {\left( {X_B^{key}|E} \right)_\omega } \ge n\log \frac{1}{{c\left( \delta  \right)}} - H_{\max }^{\epsilon}{\left( {{X_B^{key}}|{A^n}} \right)_\omega },
\label{inequality_uncertainty}
\end{equation}
where $c\left( \delta  \right)$ quantifies the overlap of the two measurements and is independent of the measured states, which considers the detectors' discretization process and has the form:
\begin{equation}\
c\left( \delta  \right) = \frac{1}{{2\pi }}{\delta ^2} \cdot S_0^{\left( 1 \right)}{\left( {1,\frac{{{\delta ^2}}}{4}} \right)^2},
\end{equation}
where $S_0^{\left( 1 \right)}$ is the ${0^{th}}$ radial prolate spheroidal wave function of the first kind~\cite{J.Math.Phys.51.072105.2010}, which can be well approximated by $c\left( \delta  \right) \approx {{{\delta ^2}} \mathord{\left/ {\vphantom {{{\delta ^2}} {\left( {2\pi } \right)}}} \right.\kern-\nulldelimiterspace} {\left( {2\pi } \right)}}$ if the interval length $\delta $ is not large. $H_{\max }^{\epsilon}{\left( {{X_B^{key}}|{A^n}} \right)_\omega }$ is the smooth max-entropy between Bob's data and Alice's system with smoothing parameter $\epsilon$. In Eq.~(\ref{inequality_uncertainty}), we assume that the random sampling of detections are i.i.d. The goal of estimating the smooth min-entropy $H_{\min }^\epsilon \left( {{X_B^{key}}|E} \right)$ is to give an upper bound of the smooth max-entropy $H_{\max }^{\epsilon}{\left( {{X_B^{key}}|{A^n}} \right)_\omega }$.

To estimate the upper bound of $H_{\max }^{\epsilon}{\left( {{X_B^{key}}|{A^n}} \right)_\omega }$, first due to the data processing inequality \cite{IEEE.Trans.Inf.Theory.56.4674.2010}, we can obtain that

\begin{equation}\
H_{\max }^{\epsilon}{\left( {{X_B^{key}}|{A^n}} \right)_\omega } \le H_{\max }^{\epsilon}{\left( {{X_B^{key}}|{X_A^{key}}} \right)_\omega },
\end{equation}
and we need to bound the correlation between data $X_B^{key}$ and $X_A^{key}$. For that we exploit the average distance,
\begin{equation}\
d\left( {X,Y} \right) = \frac{1}{n}\sum\limits_{i = 1}^n {\left| {{{ {X} }_i} - {{ Y }_i}} \right|},
\end{equation}
to give the bound of the smooth max-entropy. It has been shown in Ref.~\cite{Phys.Rev.Lett.109.100502.2012} that if $\Pr \left[ {d\left( {X_B^{key},X_A^{key}} \right) \ge d} \right] \le {\epsilon ^2}$ holds, we can always give a bound by
\begin{equation}\
H_{\max }^\epsilon \left( {{{X_B^{key}}|{X_A^{key}}}} \right)  \le  n{\log _2}\gamma \left( d \right).
\end{equation}
where $\gamma $ is a function arising from a large deviation consideration, which reads
\begin{equation}\
\gamma \left( t \right) = \left( {t + \sqrt {{t^2} + 1} } \right){\left[ {{t \mathord{\left/
				{\vphantom {t {\left( {\sqrt {{t^2} + 1}  - 1} \right)}}} \right.
				\kern-\nulldelimiterspace} {\left( {\sqrt {{t^2} + 1}  - 1} \right)}}} \right]^t}.
\end{equation}
However, we have only data $X_A^{pe}$ and $X_B^{pe}$ with $m$ length to perform the parameter estimation rather than data $X_A^{key}$ and $X_B^{key}$, thus parameter $d$ needs to be bounded by exploiting the data only consumed in parameter estimation step. Two functions need to be defined first, one is the average second moment of the difference between two sequences, which reads
\begin{equation}\
{d_2}\left( {X,Y} \right) = \frac{1}{N}\sum\limits_{k = 1}^N {{{\left| {{X^k} - {Y^k}} \right|}^2}},
\label{average_second_moment}
\end{equation}
and the other is the average second moment for the discretized data measurements, which is denoted by
\begin{equation}\
{m_2}\left( X \right) = \frac{1}{N}\sum\limits_{k = 1}^N {{{\left| {{X^k} - {\alpha  \mathord{\left/
						{\vphantom {\alpha  \delta }} \right.
						\kern-\nulldelimiterspace} \delta }} \right|}^2}}.
\label{average_second_moment_discretized_data}
\end{equation}
Then we check whether the average distance ${d^{PE}} = d\left( {X_A^{pe},X_B^{pe}} \right)$ is not larger than a certain threshold $d_0$. They continue the protocol if ${d^{PE}} \le {d_0}$ and abort the protocol otherwise. In the case of the protocol proceeding, Alice and Bob calculate the average second moments of their data respectively, which denote $V_{{X_A}}^{PE} = {m_2}\left( {X_A^{pe}} \right)$ and $V_{{X_B}}^{PE} = {m_2}\left( {X_B^{pe}} \right)$ according to Eq.~(\ref{average_second_moment_discretized_data}), and they also compute the average second moments between their data by $V_d^{PE} = {d_2}\left( {X_A^{pe},X_B^{pe}} \right)$ according to Eq.~(\ref{average_second_moment}).

With the help of Serfling's large deviation bound~\cite{Ann.Stat.2.39.1974}, we can finally bound the max-entropy by
\begin{equation}\
H_{\max }^\epsilon \left( {{{X_B^{key}}|{X_A^{key}}}} \right)  \le  n{\log _2}\gamma \left( d_{0} + \mu \right),
\end{equation}
where $\mu$ describes the statistical fluctuation deviating from $d\left( {X_B^{key},X_A^{key}} \right)$, which denotes
\begin{equation}\
\mu  = \sqrt {2{{\log }_2}{\xi ^{ - 1}}} \frac{{N{\sigma _*}}}{{m\sqrt n }} + \frac{{4\left( {{\alpha  \mathord{\left/
					{\vphantom {\alpha  \delta }} \right.
					\kern-\nulldelimiterspace} \delta }} \right){{\log }_2}{\xi ^{ - 1}}}}{3}\frac{N}{{nm}},
\end{equation}
with
\begin{align}
\sigma _*^2 =& \frac{m}{N}\left( {V_d^{PE} - \frac{m}{N}{{\left( {{d^{PE}}} \right)}^2}} \right) + \frac{m}{N}\left( {V_{{X_A}}^{PE} + V_{{X_B}}^{PE} + 2\frac{\nu }{{{\delta ^2}}}} \right)
+ 2\frac{m}{N}\sqrt {\left( {V_{{X_A}}^{PE} + \frac{\nu }{{{\delta ^2}}}} \right)\left( {V_{{X_B}}^{PE} + \frac{\nu }{{{\delta ^2}}}} \right)},
\end{align}
and
\begin{align}
\xi  =& {\left( {{\epsilon _s} - {\epsilon _1} - 2\sqrt {2n\Gamma \left( {\alpha ,T,{M_{th}}} \right)} } \right)^2}
 - 2\exp \left( { - 2{{\left( {{\nu  \mathord{\left/{\vphantom {\nu  \alpha }} \right.\kern-\nulldelimiterspace} \alpha }} \right)}^2}\frac{{n{m^2}}}{{N\left( {m + 1} \right)}}} \right).
\end{align}
$\nu$ is the smallest real number making $\xi$ positive. If there exist $\nu$ such that $\xi$ is positive and ${\epsilon _1} - 2\sqrt {2\Gamma \left( {\alpha ,T,{M_{th}}} \right)}  < {\epsilon _s}$ is satisfied, the final secret key rate under coherent attacks can be written as
\begin{equation}\
{K_{coherent}} = {{{\ell _{Low}}} \mathord{\left/{\vphantom {{{\ell _{Low}}} N}} \right.\kern-\nulldelimiterspace} N},
\end{equation}
where ${\ell _{Low}}$ is the lower bound of the secure key length, which reads
\begin{equation}\
{\ell _{Low}}  =n\left[ {\log \frac{1}{{c\left( \delta  \right)}} - \log \gamma \left( {{d_0} + \mu } \right)} \right] - {\ell _{EC}} - \log \frac{1}{{\varepsilon _1^2{\varepsilon _c}}}+2.
\label{keyrate}
\end{equation}
Otherwise, we denote that the key rate ${K_{coherent}} = 0$. The detailed proof of this section can be seen in Ref.~\cite{Phys.Rev.A.90.042325.2014}.

%%%%%%%%%%%%%%%%%%%%%%% References %%%%%%%%%%%%%%%%%%%%%%%%%


\begin{thebibliography}{99}
	
%%%%%%%%%%%%%%%%%%%%%%%%%%%%%%%%%%%%%%%%%%%%%%%%%%%%%%%%%%%%%%%%%%%%%%%%%%%%%%%%%%%%%%%%%%%%%%%%%%%%%%%%%%%%%%%%%%%%%%%%%%%%%%%%%%%%%%%%%%%%%%%%%%%%%%%%%%%%%%%%%%%%%%
\bibitem{RevModPhys.74.145}	Gisin, N., Ribordy, G., Tittel, W. \& Zbinden, H. Quantum cryptography. {\it Rev. Mod. Phys.} {\bfseries 74}, 145 (2002).

\bibitem{RevModPhys.81.1301} Scarani, V. {\it et al.} The security of practical quantum key distribution. \textit{Rev. Mod. Phys.} \textbf{81}, 1301 (2009).

\bibitem{Pirandola_RevModPhys_2019} Pirandola, S. {\it et al.} Advances in Quantum Cryptography, arXiv:1906.01645 (2019).

\bibitem{RevModPhys.84.621}	Weedbrook, C. \textit{et al.} Gaussian quantum information. \textit{Rev. Mod. Phys.} \textbf{84}, 621 (2012).

\bibitem{Entropy.17.6072}Diamanti, E. \& Leverrier, A. Distributing secret keys with quantum continuous variables: principle, security and implementations. \textit{Entropy} \textbf{17}, 6072 (2015).


\bibitem{Phys.Rev.Lett.88.057902.2002}	Grosshans, F. \& Grangier, P. Continuous variable quantum cryptography using coherent states. \textit{Phys. Rev. Lett.} \textbf{88}, 057902 (2002).

%\bibitem{Nature.421.238.2003}	F. Grosshans, G. Van Ache, J. Wenger, R. Brouri, N. J. Cerf, and P. Grangier, Nature \textbf{421}, 238 (2003).

\bibitem{Phys.Rev.Lett.93.170504.2004}	Weedbrook, C. \textit{et al.}  Quantum cryptography without switching. \textit{Phys. Rev. Lett.} \textbf{93}, 170504 (2004).


\bibitem{Nature.Photon.13.839.2019} Zhang, G. \textit{et al.} An integrated silicon photonic chip platform for continuous-variable quantum key distribution. \textit{Nature Photon.} \textbf{13}, 839 (2019).

\bibitem{Zhang_Arxiv_2020} Zhang, Y.-C., Chen, Z., Pirandola, S., Wang, X., Zhou, C., Chu, B., Zhao, Y., Xu, B., Yu, S. \& Guo, H. Long-distance continuous-variable quantum key distribution over $202.81$ {\rm km} fiber,  arXiv:2001.02555 (2020).


\bibitem{Photon.Tech.Lett.30.650.2018} Karinou, F. \textit{et al.}  Toward the integration of cv quantum key distribution in deployed optical networks. \textit{IEEE Photonics Technology Letters} \textbf{30}, 650 (2018).

\bibitem{Communications.Physics.2.9.2019} Eriksson, T. A. \textit{et al.} Wavelength division multiplexing of continuous variable quantum key distribution and 18.3 tbit/s data channels. \textit{Communications Physics} \textbf{2}, 9 (2019).

%\bibitem{New.J.Phys.12.103042.2010}	Qi, B., Zhu, W., Qian, L. \& Lo, H. K. Feasibility of quantum key distribution through a dense wavelength division multiplexing network. \textit{New J. Phys.} \textbf{12}, 103042 (2010).
%
%\bibitem{New.J.Phys.17.043027.2015}	Kumar, R., Qin H. \& Alle\'aume, R. Coexistence of continuous variable QKD with intense DWDM classical channels. \textit{New J. Phys.} \textbf{17}, 043027 (2015).

\bibitem{field_test} Zhang, Y. \textit{et al.} Continuous-variable QKD over 50 km commercial fiber. \textit{Quantum Sci. Technol.} \textbf{4}, 035006 (2019).

\bibitem{Phys.Rev.Lett.102.180504.2009}	Leverrier, A. \& Grangier, P. Unconditional security proof of long-distance continuous-variable quantum key distribution with discrete modulation. \textit{Phys. Rev. Lett.} \textbf{102}, 180504 (2009).

\bibitem{Phys.Rev.A.83.042312.2011}	Leverrier, A. \& Grangier, P. Continuous-variable quantum-key-distribution protocols with a non-Gaussian modulation. \textit{Phys. Rev. A} \textbf{83}, 042312 (2011).

\bibitem{arXiv.1805.04249}	Li, Z., Zhang, Y. \& Guo, H. User-defined quantum key distribution. arXiv:1805.04249 (2018).

\bibitem{Nat.Phys.4.726.2008} Pirandola, S., Mancini, S., Lloyd, S. \& Braunstein, S. L. Continuous-variable quantum cryptography using two-way quantum communication. \textit{Nat. Phys.} \textbf{4}, 726 (2008).

\bibitem{Int.J.Quantum.Inf.10.1250059.2012}	Sun, M., Peng, X., Shen, Y. \& Guo, H. Security of a new two-way continuous-variable quantum key distribution protocol. \textit{Int. J. Quantum Inf.} \textbf{10}, 1250059 (2012).

\bibitem{J.Phys.B.47.035501.2014}	Zhang, Y. \textit{et al.} Improvement of two-way continuous-variable quantum key distribution using optical amplifiers. \textit{J. Phys. B: At. Mol. Opt. Phys} \textbf{47}, 035501 (2014).

\bibitem{Phys.Rev.A.92.062323.2015}	Ottaviani, C., Mancini, S. \& Pirandola, S. Two-way Gaussian quantum cryptography against coherent attacks in direct reconciliation. \textit{Phys. Rev. A} \textbf{92}, 062323 (2015).

\bibitem{Sci.Rep.6.22225.2016}	Ottaviani, C. \& Pirandola, S. General immunity and superadditivity of two-way Gaussian quantum cryptography. \textit{Sci. Rep.} \textbf{6}, 22225 (2016).

\bibitem{J.Phys.B.At.Mol.Opt.Phys.50.035501.2017} Zhang, Y., Li, Z., Zhao, Y., Yu, S. \& Guo, H. Numerical simulation of the optimal two-mode attacks for two-way continuous-variable quantum cryptography in reverse reconciliation. \textit{J. Phys. B: At. Mol. Opt. Phys.} \textbf{50}, 035501 (2017).


\bibitem{New.J.Phys.20.103016.2018}	Huang, A., Barz, S., Andersson, E. \& Makarov, V. Implementation vulnerabilities in general quantum cryptography. \textit{New J. Phys.} \textbf{20}, 103016 (2018).

\bibitem{Phys.Rev.Lett.120.040406.2018} Thearle, O. \textit{et al.} Violation of Bell's inequality using continuous variable measurements.\textit{ Phys. Rev. Lett.} \textbf{120}, 040406 (2018).


\bibitem{Nat.Photon.9.397.2015}	Pirandola, S. \textit{et al.} High-rate measurement-device-independent quantum cryptography. \textit{Nat. Photon.} \textbf{9}, 397 (2015).

\bibitem{Phys.RevA.89.052301} Li, Z., Zhang, Y.-C., Xu, F., Peng, X. \& Guo, H. Continuous-variable measurement-device-independent quantum key distribution. \textit{Phys. Rev. A} \textbf{89}, 052301 (2014).

\bibitem{Phys.RevA.90.052325} Zhang, Y.-C., Li, Z., Yu, S., Gu, W., Peng, X. \& Guo, H. Continuous-variable measurement-device-independent quantum key distribution using squeezed states. \textit{Phys. Rev. A} \textbf{90}, 052325 (2014).

\bibitem{Nat.Commun.6.8795.2015} Gehring, T. \textit{et al.} Implementation of continuous-variable quantum key distribution with composable and one-sided-device-independent security against coherent attacks. \textit{Nat. Commun.} \textbf{6}, 8795 (2015).

\bibitem{Optica.3.634.2016}	Walk, N. \textit{et al.} Experimental demonstration of Gaussian protocols for one-sided device-independent quantum key distribution. \textit{Optica} 3, 634 (2016).

\bibitem{LS.JW4A.33} Zhang, Y., Li, Z., Yu, S. \& Guo, H. Composable security analysis for continuous variable measurement-device-independent quantum key distribution. \textit{Optical Society of America, Laser Science} \textbf{JW4A.33} (2017).

\bibitem{Phys.Rev.A.97.052327.2018} Lupo, C., Ottaviani, C., Papanastasiou, P. \& Pirandola, S. Continuous-variable measurement-device-independent quantum key distribution: Composable security against coherent attacks. \textit{Phys. Rev. A} \textbf{97}, 052327 (2018).

\bibitem{Phys.Rev.A.98.012314.2018}	Chen, Z., Zhang, Y., Wang, G., Li Z. \& Guo, H. Composable security analysis of continuous-variable measurement-device-independent quantum key distribution with squeezed states for coherent attacks. \textit{Phys. Rev. A} \textbf{98}, 012314 (2018).

\bibitem{Phys.Rev.Lett.109.100502.2012}	Furrer, F., \textit{et al.} Continuous variable quantum key distribution: finite-key analysis of composable security against coherent attacks. \textit{Phys. Rev. Lett.} \textbf{109}, 100502 (2012).

\bibitem{Phys.Rev.A.90.042325.2014}	Furrer, F. Reverse-reconciliation continuous-variable quantum key distribution based on the uncertainty principle. \textit{Phys. Rev. A} \textbf{90}, 042325 (2014).

\bibitem{Phys.Rev.A.87.022308.2013}  Weedbrook, C. Continuous-variable quantum key distribution with entanglement in the middle. \textit{Phys. Rev. A} \textbf{87}, 022308 (2013).


%constraints in two modes attacks
\bibitem{New.J.Phys.15.113046.2013}	Pirandola, S. Entanglement reactivation in separable environments. \textit{New J. Phys.} \textbf{15}, 113046 (2013).

\bibitem{Phys.Rev.A.91.022320.2015}	Ottaviani, C., Spedalieri, G., Braunstein,S. L. \& Pirandola, S. Continuous-variable quantum cryptography with an untrusted relay: Detailed security analysis of the symmetric configuration. \textit{Phys. Rev. A} \textbf{91}, 022320 (2015).

\bibitem{arXiv:1509.04144} Ottaviani, C., Spedalieri, G., Braunstein, S. L. \& Pirandola, S. CV-MDI-QKD: One-mode Gaussian attacks are not enough. arXiv:1509.04144 (2015).

%D-W relation
\bibitem{Proc.Roy.Soc.A.461.207.2005} Devetak, I., \& Winter, a. Distillation of secret key and entanglement from quantum states. \textit{Proc. Roy. Soc. A} \textbf{461}, 207 (2005).

%holevo bound
\bibitem{Probl.Inf.Transm.9.177.1973} Holevo, A. S. Bounds for the quantity of information transmitted by a quantum communication channel. \textit{Probl. Inf. Transm.} \textbf{9}, 177 (1973).

%Gaussian state extramelity theorem
\bibitem{Phys.Rev.Lett.97.190503.2006} Garc\'ia-Patr\'on, R. \& Cerf, N. J. Unconditional optimality of Gaussian attacks against continuous-variable quantum key distribution. \textit{Phys. Rev. Lett.} \textbf{97}, 190503 (2006).

\bibitem{Phys.Rev.Lett.96.080502.2006} Wolf, M. M., Giedke, G. \& Cirac, J. I. Extremality of Gaussian quantum states. \textit{Phys. Rev. Lett.} \textbf{96}, 080502 (2006).

\bibitem{Phys.Rev.Lett.114.070501.2015} Leverrier, A. Composable security proof for continuous-variable quantum key distribution with coherent states. \textit{Phys. Rev. Lett.} \textbf{114}, 070501 (2015).

\bibitem{Phys.Rev.Lett.118.200501.2017} Leverrier, A. Security of continuous-variable quantum key distribution via a Gaussian de Finetti reduction. \textit{Phys. Rev. Lett.} \textbf{118}, 200501 (2017).

\bibitem{Entropy.21.652.2019} Chen, Z., Zhang, Y., Wang, X., Yu, S. \& Guo, H. Improving parameter estimation of entropic uncertainty relation in continuous-variable quantum key distribution. \textit{Entropy} \textbf{21}, 652 (2019).

\bibitem{Nat.Commun.8.15043.2017} Pirandola, S., Laurenza, R., Ottaviani, C. \&  Banchi, L. Fundamental limits of repeaterless quantum communications. Nat. Commun. \textbf{8}, 15043 (2017).

\bibitem{Phys.Rev.Lett.102.130501.2009}	Garc\'ia-Patr\'on, R. \& Cerf, N. J. Continuous-variable quantum key distribution protocols over noisy channels. Phys. Rev. Lett. \textbf{102}, 130501 (2009).

\bibitem{arXiv.0512258.2006} Renner, R. Security of quantum key distribution. Ph.D. thesis, Swiss Federal Institute of Technology (ETH) Zurich, 2006; arXiv:quant-ph/0512258.

\bibitem{New.J.Phys.15.053049.2013} Eberle, T. \textit{et al.} Gaussian entanglement for quantum key distribution from a single-mode squeezing source. \textit{New J. Phys.} \textbf{15}, 053049 (2013).

\bibitem{Nat.Phys.6.659.2010} Berta, M., Christandl, M., Colbeck, R., Renes, J. M. \& Renner, R. The uncertainty principle in the presence of quantum memory.\textit{ Nat. Phys.} \textbf{6}, 659 (2010).

\bibitem{Phys.Rev.Lett.106.110506.2011} M. Tomamichel, R. Renner, Phys. Rev. Lett. \textbf{106}, 110506 (2011).

\bibitem{Commun.Math.Phys.306.165.2011} Furrer, F., {\AA}berg, J., Renner, R. Min-and max-entropy in infinite dimensions. \textit{Commun. Math. Phys.} \textbf{306}, 165 (2011).

\bibitem{J.Math.Phys.55.122205.2014} Furrer, F., Berta, M., Tomamichel, M., Scholz, V. B. \& Christandl, M. Position-momentum uncertainty relations in the presence of quantum memory. \textit{J. Math. Phys.} \textbf{55}, 122205 (2014).

\bibitem{J.Math.Phys.51.072105.2010} Kiukas, J. \& Werner, R. F. Maximal violation of Bell inequalities by position measurements. \textit{J. Math. Phys.} \textbf{51}, 072105 (2010).

\bibitem{IEEE.Trans.Inf.Theory.56.4674.2010} Tomamichel, M., Colbeck, R. \& Renner, R. Duality between smooth min-and max-entropies. \textit{IEEE Trans. Inf. Theory} \textbf{56}, 4674 (2010).

\bibitem{Ann.Stat.2.39.1974} Serfling, R. J. Probability inequalities for the sum in sampling without replacement. \textit{Ann. Stat.} \textbf{2}, 39 (1974).





\end{thebibliography}
\end{document}